\documentclass[prd, twocolumn, nofootinbib, floatfix]{revtex4-1}
\usepackage{amsmath}
\usepackage{graphicx}
\usepackage{dcolumn}
\usepackage{bm}
\usepackage{epsfig}
\usepackage{amssymb,latexsym,mathrsfs}
\usepackage{graphicx}
\usepackage{color}
\usepackage{hyperref}

\usepackage{float}

\hypersetup{
    colorlinks=true,
    linkcolor=red,
    citecolor=blue,
}

\newcommand{\be}{\begin{equation}}
\newcommand{\ee}{\end{equation}}
\newcommand{\bs}{\begin{split}} 
\newcommand{\bea}{\begin{eqnarray}}
\newcommand{\eea}{\end{eqnarray}}

\newcommand{\alm}{\alpha_M} 
\newcommand{\alb}{\alpha_B} 
\newcommand{\ms}{M_\star^2}
\newcommand{\gm}{G_{\rm matter}} 
\newcommand{\gl}{G_{\rm light}}

\begin{document}

\title{Gravitational Wave Distances in Horndeski Cosmology} 

\author{Eric V.\ Linder$^{1,2}$} 
\affiliation{${}^1$Berkeley Center for Cosmological Physics \& Berkeley Lab, 
University of California, Berkeley, CA 94720, USA\\
${}^2$Energetic Cosmos Laboratory, Nazarbayev University, 
Nur-Sultan 010000, Kazakhstan
}

\begin{abstract} 
Gravitational wave propagation encounters a spacetime friction from 
a running Planck mass in modified gravity, causing the luminosity 
distance to deviate from that in general relativity (or given by 
the photon luminosity distance to the source), thus making it a 
valuable cosmological probe. We present the exact expression  
for the cosmological distance deviation in Horndeski gravity including theories that have a 
$G_5$ term yet propagate at the speed of light. An especially 
simple result ensues for coupled Gauss-Bonnet gravity, which we 
use to show it does not give a viable cosmology. We also generalize 
such coupling, and review the important connection of gravitational 
wave cosmological distance deviations to growth of cosmic structure measured by 
redshift space distortions. 
\end{abstract}

\date{\today} 

\maketitle

\section{Introduction}

Gravitational waves (GW) provide a new probe of cosmology as well as  
of gravitation in the strong field regime. For cosmology, standard 
siren distances from GW are a new type of distance measure, crosschecking 
those from photon luminosity distance (e.g.\ Type Ia supernovae) and 
angular distance (e.g.\ baryon acoustic oscillations). Thus they can map 
out the cosmic expansion history. However, the GW distance depends on the 
propagation of gravitational waves -- if this differs from that in general 
relativity then one must account for this in interpretation of the distance.  

GW propagation can differ from general relativity in its speed of 
propagation and an additional friction beyond the Hubble friction of 
expanding space. This new spacetime friction is due to a change in the  
coupling of gravity to spacetime curvature, and can be thought of as 
an evolving gravitational strength (Newton's constant) or running Planck 
mass. (Additional effects such as graviton mass and source terms can 
enter, but do not in the Horndeski class of gravity we consider here.)  

Thus GW not only map out cosmic expansion history, but cosmic gravity  
history, and we will highlight as well a connection to cosmic growth 
history. Thus a simple expression for the deviation of GW propagation 
from general relativity is of interest. This is constrained by the 
implication of near simultaneous GW and electromagnetic bursts from 
GW170817 \cite{gwobs} indicating that the speed of GW propagation 
equals the speed of light, within the most direct interpretation (see 
\cite{battye,melville} for other possibilities). This stringently 
restricts the gravitation theory. 

Working within the class of Horndeski theory, the most general 
scalar-tensor theory with second order equations of motion, the 
restriction to the speed of light is usually taken to remove one of the four terms in 
the action and prevent another from depending on the scalar field 
motion. This is not absolute, however, and can lead to some interesting 
cases. 

In Section~\ref{sec:dlgw} we 
show how the general expression for the GW distance deviation looks 
in the usual interpretation and in the extended one. Section~\ref{sec:cgb} 
treats a special case of the latter situation, involving a coupling to 
the Gauss-Bonnet term, of particular interest since it is a geometric 
invariant. In Section~\ref{sec:gencoup} and \ref{sec:g50} we examine 
a generalization, and another special case, respectively. 
The extraordinary connection between 
GW propagation and cosmic structure growth deviations from general 
relativity in some theories is visited in Sec.~\ref{sec:grow} and 
we conclude in Sec.~\ref{sec:concl}.

\section{GW Distance Deviation} \label{sec:dlgw} 

The propagation equation for the GW amplitude $h$ in a cosmological background 
is \cite{0004156,1406.7139,lomtay,nis} 
\be 
\ddot h + (3+\alm)H\dot h+k^2h=0\,, \label{eq:gwprop} 
\ee 
where an overdot is a derivative with respect to time, 
$\alm=d\ln\ms/d\ln a$ is the Planck mass evolution rate, $H=\dot a/a$ is the 
Hubble parameter, $a$ the cosmic scale factor, and $k$ the wavenumber. 
As stated, we work within the Horndeski class of gravity and have 
already set the GW speed of propagation to the speed of light (i.e.\ $c_T=1$). 

The emitted amplitude is predicted by general relativity based on 
the detected GW characteristics (and it is assumed general relativity 
holds in the emission process, as most viable cosmic gravity theories 
have screening mechanisms that restore to general relativity in regions 
of much higher density than the cosmic background). Comparing this to 
the observed amplitude gives a GW distance to the source through the 
cosmic inverse square distance law 
(energy $\sim$ amplitude$^2$  $\sim$ 1/distance$^2$ so $d_L\sim 1/h$). 

A clear derivation of the solution to the GW propagation Eq.~\eqref{eq:gwprop} 
was given by \cite{nis,1711.03776}, and related directly to GW distances for 
Horndeski gravity by 
\cite{1712.08623,nsg}: 
\be 
d_{L,GW}(a)=d_L^{GR}(a)\,\left[\frac{\ms(a=1)}{\ms(a)}\right]^{1/2}\,. 
\label{eq:devms} 
\ee 
Thus the key quantity of interest from modified gravity is the Planck 
mass evolution. 

In a general Horndeski theory, 
\be 
\ms=2\left(G_4-2XG_{4X}+XG_{5\phi}-H\dot\phi XG_{5X}\right)\,, \label{eq:msg5}
\ee 
where $G_4(\phi,X)$ and $G_5(\phi,X)$ are two of the Horndeski 
Lagrangian terms (the others do not enter GW propagation), $X=\dot\phi^2/2$, 
$\phi$ is a scalar field, and subscripts $\phi$ or $X$ denote derivatives 
with respect to that quantity \cite{belsaw}. 
Conventionally when the GW propagation speed $c_T$ is the speed of light, 
$G_5=0$ and $G_{4X}=0$, leaving just $\ms=2G_4(\phi)$. For general relativity, 
$G_4=M^2_{pl}/2$ so $\ms=M^2_{pl}$, a constant; we will work in units 
such that $M^2_{pl}=1$. 

Thus in the conventional case, 
\be
d_{L,GW}(a)=d_L^{GR}(a)\,\left[\frac{G_4(\phi(a=1))}{G_4(\phi(a))}\right]^{1/2}\,. \label{eq:devg4std} 
\ee
However, one can also have the GW propagation speed $c_T\equiv 1+\alpha_T$ equal to the speed 
of light when 
\be 
\alpha_T\sim 2G_{4X}-2G_{5\phi}-(\ddot\phi-H\dot\phi)\,G_{5X}=0\,. \label{eq:at}
\ee 
Substituting this into Eq.~\eqref{eq:msg5} gives 
\be 
\ms=2G_4(\phi,X)-\dot\phi \dot G_5(\phi,X)\,, \label{eq:ms4c1} 
\ee 
where $\dot G_i=\dot\phi G_{i\phi}+\dot X G_{iX}=\dot\phi G_{i\phi}+\dot\phi\ddot\phi G_{iX}$. 

The general expression for distance deviations then becomes 
\be
d_{L,GW}(a)=d_L^{GR}(a)\,\left(\frac{[2G_4-\dot\phi\dot G_5](a=1)}{[2G_4-\dot\phi\dot G_5](a)}\right)^{1/2}\,. \label{eq:devg5} 
\ee
Given a Horndeski theory, one solves the equations of motion (including 
for $\phi(a)$), and can determine the GW distance deviation from general 
relativity.

\section{Coupled Gauss-Bonnet Gravity} \label{sec:cgb} 

For certain theories within the Horndeski class, the above expressions 
work out particularly simply. An interesting case is coupled 
Gauss-Bonnet gravity, demonstrated to be part of the Horndeski class 
in \cite{1105.5723}. The Gauss-Bonnet invariant 
\be 
\mathcal{G}=R^2-4R_{ab}R^{ab}+R_{abcd}R^{abcd}\,, 
\ee 
is a particular scalar combination 
of the Riemann, Ricci tensor, and Ricci scalar curvatures, and so an 
important geometric object. Being a topological term, by itself it does 
not alter the equations of motion from general relativity, however promoting 
it to a function $f(\mathcal{G})$ or coupling it to a scalar field as 
$f(\phi)\mathcal{G}$ in the action gives it dynamics and hence modifies gravity. 

We consider the latter case, with the action being the usual Ricci scalar 
plus a $f(\phi)\mathcal{G}$ term. There can be kinetic and potential 
terms of the scalar field as well, but they will not affect GW propagation. 
As coupled Gauss-Bonnet (CGB) gravity possesses a $G_5$ term and a $G_4$ 
term depending on $X$, normally it has a GW speed of propagation different 
from the speed of light. However, \cite{1908.07555,2003.13724} 
demonstrated the condition such that a restricted case survives: the 
CGB speed of GW propagation will be the speed of light if the coupling 
satisfies 
\be 
\ddot f=H\dot f\,. \label{eq:fdot} 
\ee 

If we then evaluate Eq.~\eqref{eq:msg5} or \eqref{eq:ms4c1} for the CGB terms 
\be 
G_4=\frac{1}{2}+4f_{\phi\phi}X(2-\ln X)\,,\ \  G_5=-4f_\phi\ln X\,, 
\ee 
then we obtain 
\be 
\ms=1+8H\dot f=1+8H\dot\phi f_\phi\,. 
\ee 
Eq.~\eqref{eq:fdot} can be readily solved to give $\dot f=ca$, where 
$c$ is a constant. Note that we never need to solve the scalar field 
equation of motion -- this is a model independent result! (If we are 
given $f(\phi)$, then we can find $\phi(a)$ through $f_\phi\dot\phi=ca$.) 

Thus, independent of the specific coupling, 
\be 
d_{L,GW}(a)=d_L^{GR}(a)\,\left(\frac{1+8cH_0}{1+8caH}\right)^{1/2}\,. 
\quad {\rm (CGB)} 
\ee 
Given a cosmological background expansion, i.e.\ Hubble parameter $H(a)$, 
which may depend on the coupling form, as well as kinetic and potential 
terms, one has an exact prediction for the GW distance deviation. 

Note that since $\ms=1+8caH=1+8c\dot a$, then $\ms$ blows up as $a\to0$. 
We explore this further below. 
CGB is also sometimes used as an inflation 
theory, and note that for $H$ constant the solution to Eq.~\eqref{eq:fdot} 
is $\dot f=ce^{H_i t}$ and $\ms=1+8cH_ie^{H_i t}$. 

Let us examine some of the other gravity theory quantities, such as 
the Planck mass evolution rate $\alm$. We have 
\be 
\alm\equiv\frac{\dot\ms}{H\ms}=\frac{8c\ddot a}{H(1+8c\dot a)}\,. 
\ee  
Note that 
\be 
\alm \to \frac{a\ddot a}{\dot a^2}=-q\,, \qquad (a\to0) 
\ee 
where $q$ is the cosmic deceleration parameter. Thus, in CGB $\alm$ 
directly measures the acceleration of the universe. (And since we do 
not have $\alm\to0$, general relativity is never fully restored in 
the early universe.) For the inflation case we indeed find $\alm=1$. 

Since in our universe $\ddot a$ changes sign as the expansion moves 
from matter domination to our present accelerated epoch, then there 
will be a time when $\alm$ crosses zero. Recall that $\alm=0$ is 
referred to as No Run Gravity \cite{nrg} (when it is a persistent 
condition) and gives zero gravitational 
slip. That is an interesting, if momentary, property of CGB. 

The property functions such as $\alm$ and $\alpha_T$ give a useful 
perspective on observational effects of modified gravity \cite{belsaw}. 
Computing the function $\alpha_B$ that describes the braiding, or 
mixing between the scalar kinetic term and the metric, 
\be 
\alpha_B=\frac{-8H\dot f}{\ms}=\frac{-8caH}{1+8caH}\,. 
\ee 
We see that this has $\alpha_B\to-1$ at early times, and this holds 
for the inflationary solution as well. This means that in early 
matter domination $\alpha_B=2\alm=-1$, in radiation domination 
$\alpha_B=\alm=-1$, and in inflation $\alpha_B=-\alm=-1$. The last 
relation is characteristic of $f(R)$ gravity as well. The remaining 
property function $\alpha_K$, the kineticity, does not have much 
observable impact, affecting the spatial clustering of the scalar field. 
We find it depends on $f_{\phi\phi\phi\phi}$, and hence is not 
model independent. 

The gravitational strength, relative to Newton's constant in general 
relativity, entering the growth of structure is denoted $\gm$, and 
that for light propagation is $\gl$; the difference between them is 
referred to as the gravitational slip. Both involve combinations of 
$\alm$ and $\alpha_B$; their expressions can be found in, for 
example, \cite{lmg}. Evaluating for CGB, 
\bea 
\gm&=&\frac{2H\ddot a-H^2\dot a-8c(H^2\dot a^2-2\ddot a^2)}{H(1+8c\dot a)[2\ddot a-H\dot a-4c(3H\dot a^2-4\dot a\ddot a)]}\\ 
\gl&=&\frac{2H\ddot a-H^2\dot a+4cH\dot a\ddot a-8c(H^2\dot a^2-\ddot a^2)}{H(1+8c\dot a)[2\ddot a-H\dot a-4c(3H\dot a^2-4\dot a\ddot a)]}\,. 
\eea 
Note that when $c\to0$, and the coupling vanishes, 
then $\gm\to1$, $\gl\to1$, i.e.\ we recover 
general relativity. 

In the matter dominated or radiation dominated eras (or any with background 
equation of state $w_b>-1/3$), $\gm$ and $\gl$ approach zero going into the 
past. This is not surprising since we found that $\alm$ and $\alb$ go  
to constants, and $\ms$ in the denominator blows up. This vanishing of 
gravity does not make for a viable cosmology. If the dominant component has 
$w_b<-1/3$ then $\gm$ and $\gl$ approach one going into the past. For 
inflation, with $w_b=-1$, going back into the past $\gm$ and $\gl$ will 
be one as in general relativity, but when inflation lasts more than a 
few e-folds, i.e.\ $c\,e^{H_i t}$ gets large, then again gravity vanishes. 

Thus there is no valid inflation nor late time cosmology for coupled 
Gauss-Bonnet gravity with GW propagation at the speed of light, 
independent of the model, i.e.\ coupling function. (One could use it 
for inflation with GW speed $c_T\ne1$, but the gravitation theory 
must somehow change by the late universe.)

\section{Generalizing the Coupling} \label{sec:gencoup} 

While coupled Gauss-Bonnet gravity has some attractive features, 
such as use of the geometric invariant, it did not give rise to a 
viable cosmology with GW speed $c_T=1$. Let us explore whether 
we can keep some useful features to find the GW distance deviation 
in a viable theory. In setting $\alpha_T=0$, CGB led to the constraint 
on the coupling $\ddot f=H\dot f$, which has the advantage of a 
model independent form $\dot f=ca$, and one does not have to 
know the dependence $f(\phi)$ to compute the GW distance 
deviation and property functions. 

We can achieve this generally by writing 
\be 
G_4=f_{\phi\phi}(\phi)\,g_4(X)\,,\qquad G_5=f_\phi(\phi)\,g_5(X)\,. 
\ee 
One obtains $\alpha_T=0$ with $\ddot f=H\dot f$ when 
\be 
g_5(X)=-X\int dx\,\frac{g_{4x}}{x^2}\ . \label{eq:g5int} 
\ee  
Solutions include 
\bea 
g_5&=&c+d\ln X\,,\  \  g_4=X\left[(c-d-1)+d\ln X\right] \label{eq:gencgb} \\ 
g_5&=&cX^n\,,\qquad\ \ \  g_4=\frac{c(1-n)}{1+n}\,X^{n+1}\\ 
g_5&=&cX^{-1}\,,\qquad\ \, g_4=2c\,\ln X\,, 
\eea  
leading respectively to  
\bea 
\ms&=&-2dH\dot f+2(d-1)Xf_{\phi\phi}\,, \label{eq:g5ln}\\ 
\ms&=&-2ncX^n H\dot f +2[n^2c/(1+n)]X^{n+1}f_{\phi\phi}\\ 
\ms&=&2cX^{-1}H\dot f+2c(-3+2\ln X)f_{\phi\phi}\,. 
\eea 

One could proceed with all the observables for these theories, 
however the presence of $f_{\phi\phi}$ in $\ms$ means that we 
have lost model independence. Only in the first case can we remove 
$f_{\phi\phi}$, by choosing $d=1$. Eqs.~\eqref{eq:gencgb} and 
\eqref{eq:g5ln} generalize 
the CGB case, which has $c=0$ (and note that 
$f_{\rm here}=-4f_{\rm CGB}$), but note this makes no difference 
for $\ms$. On the other hand, the $f_{\phi\phi}$ term in $\ms$ 
offers the hope that $\ms$ will not blow up in the past, allowing 
for a viable cosmology. Since one would have to compute this  
model by model, we do not pursue it further. 

There is another method for removing $f_{\phi\phi}$ from 
appearing in $\ms$. One takes both the condition 
$g_{4X}=g_5-Xg_{5X}$ that led to Eq.~\eqref{eq:g5int}, 
and a further condition $g_5=2g_{4X}-X^{-1}g_4$. This then 
gives 
\be 
\ms=2H\dot f\,\left(\frac{-g_4}{X}+g_{4X}-2Xg_{4XX}\right)\,. 
\ee 
Having $X$ in $\ms$ is also model independent so we 
want to remove it as well. This is accomplished with 
\be 
g_5=c+2d+d\ln X,\ \  g_4=X(c+d\ln X),\ \  \ms=-2dH\dot f\,. 
\ee 
This will have the same problems of $\ms$ blowing 
up in the past  as CGB (which is a special case with 
$c=-2$, $d=1$). 

Thus, the model independent cases we have 
examined are not viable, and the potentially viable gravity 
theories are not independent of the form of the $\phi$ 
dependence. The next section explores a middle path.

\section{$G_{5X}=0$} \label{sec:g50} 

In Eq.~\eqref{eq:at} for $\alpha_T$, a term with $\ddot\phi$ 
appears explicitly. We can remove this by setting $G_{5X}=0$, 
in the hopes of avoiding having to be explicit about the scalar 
field evolution. To keep the GW speed $c_T=1$, this then 
implies $G_{4X}=G_{5\phi}$. That in turn gives 
\be 
\ms=2(G_4-XG_{4X})\,. 
\ee 
When forming $\alm$, we will end up with a term involving 
$\dot XG_{4XX}$, which again introduces $\ddot\phi$, so we  
also take $G_{4XX}=0$. 

This implies 
\be 
G_4=\frac{1}{2}+Xf_\phi+g(\phi)\,,\quad G_5=f(\phi)+c\,, 
\ee 
yielding the simple expressions 
\bea 
\ms&=&1+2g(\phi)\\ 
\alm&=&\frac{2\dot\phi g_\phi}{H[1+2g(\phi)]}\\  
\alb&=&\frac{2\dot\phi\,(XG_{3X}-3Xf_{\phi\phi}-g_\phi)}{H[1+2g(\phi)]}\,. 
\eea 
Gravitational wave distances go as 
\be 
d_{L,GW}(a)=d_L^{GR}(a)\,\left[\frac{1+2g(a=1)}{1+2g(a)}\right]^{1/2}\,. 
\ee 
It is interesting to note that $G_5$ does not affect them, i.e.\ 
$f(\phi)$ does not enter (though  it appears in $\alb$). 
Note that $f=0$ gives a theory with 
simple scalar coupling $g(\phi)$ to 
the Ricci scalar, equivalent to $f(R)$ gravity. 

While model dependent, this class of theories at least has the 
virtue of simplicity. In the early universe, we want general 
relativity to describe cosmology so $\ms\to1$, 
implying $g\to0$ (or a constant) there, 
and $\alm$, $\alb\to0$ 
(neglecting $G_{3X}$) if $g$ and $f_\phi$ start out slowly rolling. 
At late times, if the cosmology goes 
to a frozen de Sitter state, then the numerators of $\alm$ and 
$\alb$, which involve $\dot g=\dot\phi g_\phi$ and $\dot f_\phi$, 
vanish, again restoring general relativity.

\section{Connecting GW Distances to Cosmic Growth} \label{sec:grow} 

The relation between GW distance deviations and 
growth of cosmic structure deviations from general relativity is 
an intriguing connection,  
developed in \cite{nsg} and elaborated in \cite{lmg}. There it was 
pointed out that in Horndeski theories where $\alb$ is a function 
of $\alm$, then $\gm$ and $\ms$, and hence $d_{L,GW}$ are  
connected. Not all theories do have a relation $\alb(\alm)$ since 
$G_3$, $G_4$, and $G_5$ are generally independent functions, 
even if we impose $c_T=0$. 

However the class of No Slip 
Gravity has the very direct $\gm=1/\ms$ (as does the 
non-Horndeski nonlocal gravity model in \cite{belg}). 
For the class of standard scalar-tensor theories, 
$\gl=1/\ms$. 
Another 
interesting case is Only Run Gravity \cite{lmg}, where there is 
no braiding ($\alb=0$). In that theory 
\be 
\gm=\frac{\ms+(\ms)'}{(\ms)^2}\ , 
\ee  
where a prime denotes $d/d\ln a$. 
Illustrations 
of the connection between the cosmic structure redshift space 
distortion quantity $f\sigma_8(a)$, basically the growth rate, and 
$d_{L,GW}/d_L^{GR}$ are shown in \cite{lmg} for several 
theories. 

Suppose we ask the inverse question, whether there could 
be no deviation in GW distances, yet deviation in cosmic 
growth, and vice versa. If $d_{L,GW}=d_L^{GR}$ then 
$\ms=1$ and $\alm=0$. This is called the class of No Run 
Gravity \cite{nrg}. Eq.~\eqref{eq:ms4c1} indicates this can 
occur if either there is $G_5=0$, $G_4=1/2$, or a balance 
such that 
\be 
G_4=\frac{1}{2}\left[1+\dot\phi\dot G_5\right]=\frac{1}{2}+XG_{5\phi}+X\ddot\phi G_{5X}\,.  
\ee 
This imposes a constraint on the scalar field evolution so 
it is not very generic. Regardless, since $\gm$ also 
depends on $\alb$, which can involve $G_3$, we can 
indeed have deviations in growth. 

If growth does not deviate from general relativity, this is 
called the class of Only Light Gravity \cite{lmg}. Then 
there is a differential equation relating $\alb$ and $\alm$. 
This does give a deviation in $d_{L,GW}$ that depends 
on the form adopted for $\alm$ (or equivalently $G_4$ and 
$G_5$). 

Thus, GW distance deviations and cosmic growth deviations 
do serve as complementary probes in general, while in a 
few interesting classes of gravity they can be critical  
crosschecks on each other.

\section{Conclusions} \label{sec:concl} 

Gravitational wave propagation has already had a dramatic 
impact on cosmological  gravitation theory, severely restricting 
models that do not predict propagation at the speed of light. 
The GW distance deviation from general relativity maps the 
Planck mass evolution; in the simplest interpretation it traces 
out the gravitation history of the universe. 
 
There are ways around the usual interpretation within Horndeski 
gravity that $c_T=1$ implies $G_5=0$, $G_{4X}=0$. We give the 
general expression for the GW distance deviation that does not 
have this, yet preserves $c_T=1$. A particular interesting example 
involves coupling to the Gauss-Bonnet geometric invariant, 
but we demonstrate a no go theorem that coupled Gauss-Bonnet 
gravity cannot have $c_T=1$ and give a viable description of 
our cosmology, regardless of the exact coupling. 

We extend this exploration to further forms of $G_4$ and $G_5$ 
that obey $c_T=1$, giving the forms for the GW distance 
deviation. Some have attractive properties in being model 
independent, but are not viable, while others are viable, but 
one must treat model by model. 

Other probes of cosmological gravity such as the growth rate 
of structure and light propagation can be connected to GW 
distance deviation. We show that in some cases this is a  
direct relation, hence an important crosscheck that deviations 
from general relativity are real, against systematic effects; 
in other cases the probes are complementary, working 
together to reveal the underlying nature of gravitation and 
the gravity history of our universe.

\acknowledgments 

This work is supported in part by the Energetic Cosmos Laboratory and by the 
U.S.\ Department of Energy, Office of Science, Office of High Energy 
Physics, under contract no.~DE-AC02-05CH11231.


\end{document}